%% file: 0archive-position-paper-WAFVR-hpcmaspa.tex
\documentclass[conference,onecolumn]{IEEEtran}
\IEEEoverridecommandlockouts

\usepackage{cite}
\usepackage{amsmath,amssymb,amsfonts}
\usepackage{algorithmic}
\usepackage{graphicx}
\usepackage{textcomp}
\usepackage{xcolor}
\usepackage{url}
\usepackage{enumerate}
\usepackage[normalem]{ulem}
\usepackage{caption}
\usepackage{subcaption}
\newcommand{\mape}[1]{{\textit{MAPE}#1}}
\newcommand{\mapek}[1]{{\textit{MAPE-K}#1}}
\newcommand{\monitor}[1]{{\textit{Monitor}#1}}
\newcommand{\analyze}[1]{{\textit{Analyze}#1}}
\newcommand{\plan}[1]{{\textit{Plan}#1}}
\newcommand{\execute}[1]{{\textit{Execute}#1}}
\newcommand{\knowledge}[1]{{\textit{Knowledge}#1}}

\newcommand{\Maintenance}[1]{{\emph{Maintenance}#1}}
\newcommand{\IOQoS}[1]{{\emph{I/O QoS}#1}}
\newcommand{\OST}[1]{{\emph{OST}#1}}
\newcommand{\Misconfiguration}[1]{{\emph{Misconfiguration}#1}}
\newcommand{\Scheduler}[1]{{\emph{Scheduler}#1}}

\def\BibTeX{{\rm B\kern-.05em{\sc i\kern-.025em b}\kern-.08em
    T\kern-.1667em\lower.7ex\hbox{E}\kern-.125emX}}
\begin{document}

\title{Autonomy Loops for Monitoring, Operational Data Analytics, Feedback, and Response\\ in HPC Operations}

\author{
    \IEEEauthorblockN{
    Francieli Boito\IEEEauthorrefmark{6},
    Jim Brandt\IEEEauthorrefmark{2},
    Valeria Cardellini\IEEEauthorrefmark{10},
    Philip Carns\IEEEauthorrefmark{1},
    Florina M. Ciorba\IEEEauthorrefmark{3},\\
    Hilary Egan\IEEEauthorrefmark{18},
    Ahmed Eleliemy\IEEEauthorrefmark{3},
    Ann Gentile\IEEEauthorrefmark{2},
    Thomas Gruber\IEEEauthorrefmark{16},
    Jeff Hanson\IEEEauthorrefmark{14},
    Utz-Uwe Haus\IEEEauthorrefmark{21},\\
    Kevin Huck\IEEEauthorrefmark{4},
    Thomas Ilsche\IEEEauthorrefmark{8},
    Thomas Jakobsche\IEEEauthorrefmark{3},
    Terry Jones\IEEEauthorrefmark{13},
    Sven Karlsson\IEEEauthorrefmark{7},
    Abdullah Mueen\IEEEauthorrefmark{17},\\
    Michael Ott\IEEEauthorrefmark{12},
    Tapasya Patki\IEEEauthorrefmark{19},
    Ivy Peng\IEEEauthorrefmark{20},
    Krishnan Raghavan\IEEEauthorrefmark{1},
    Stephen Simms\IEEEauthorrefmark{5},
    Kathleen Shoga\IEEEauthorrefmark{19},\\
    Michael Showerman\IEEEauthorrefmark{15},
    Devesh Tiwari\IEEEauthorrefmark{11},
    Torsten Wilde\IEEEauthorrefmark{14},
    and Keiji Yamamoto\IEEEauthorrefmark{9}}

\IEEEauthorblockA{
\IEEEauthorrefmark{6}University of Bordeaux, CNRS, Bordeaux INP, INRIA, LaBRI, Talence, FR;\\
\IEEEauthorrefmark{2}Sandia National Laboratories, CA and NM, US;
\IEEEauthorrefmark{10}University of Rome Tor Vergata, IT;\\
\IEEEauthorrefmark{1}Argonne National Laboratory, IL, US;
\IEEEauthorrefmark{3}University of Basel, Basel, CH;\\
\IEEEauthorrefmark{18}National Renewable Energy Laboratory, CO, US;\\
\IEEEauthorrefmark{16}Erlangen National High Performance Computing Center (NHR@FAU), DE;\\
\IEEEauthorrefmark{14}Hewlett Packard Enterprise, DE and US;
\IEEEauthorrefmark{21}Hewlett Packard Labs, EMEA Research Lab, CH;\\
\IEEEauthorrefmark{4}University of Oregon, Eugene, OR, US;
\IEEEauthorrefmark{8}Technische Universität Dresden, DE;\\
\IEEEauthorrefmark{13}Oak Ridge National Laboratory, TN, US;
\IEEEauthorrefmark{7}Technical University of Denmark, DK;\\
\IEEEauthorrefmark{17}University of New Mexico, NM, US;
\IEEEauthorrefmark{12}Leibniz Supercomputing Centre, DE;\\
\IEEEauthorrefmark{5}Lawrence Berkeley National Laboratory, CA, US;
\IEEEauthorrefmark{20}KTH Royal Institute of Technology, SE;\\
\IEEEauthorrefmark{19}Lawrence Livermore National Laboratory, CA, US;
\IEEEauthorrefmark{15}University of Illinois, IL, US;\\
\IEEEauthorrefmark{11}Northeastern University, MA, US;
and
\IEEEauthorrefmark{9}RIKEN R-CCS, JP}
Email:
\IEEEauthorrefmark{6}
{francieli.zanon-boito@u-bordeaux.fr},
\IEEEauthorrefmark{2}
{\{brandt, gentile\}@sandia.gov},
\IEEEauthorrefmark{10}
{cardellini@ing.uniroma2.it},\\
\IEEEauthorrefmark{1}
{\{carns, kraghavan\}@mcs.anl.gov},
\IEEEauthorrefmark{3}
{\{florina.ciorba, ahmed.eleliemy, thomas.jakobsche\}@unibas.ch},\\
\IEEEauthorrefmark{18}
{hilary.egan@nrel.gov},
\IEEEauthorrefmark{16}
{thomas.gruber@fau.de},
\IEEEauthorrefmark{14}
{\{jeff.hanson, torsten.wilde\}@hpe.com},
\IEEEauthorrefmark{21}
{uhaus@hpe.com},\\
\IEEEauthorrefmark{4}
{khuck@cs.uoregon.edu},
\IEEEauthorrefmark{8}
{thomas.ilsche@tu-dresden.de},
\IEEEauthorrefmark{13}
{trj@ornl.gov},
\IEEEauthorrefmark{7}
{svea@dtu.dk},
\IEEEauthorrefmark{17}
{mueen@unm.edu},\\
\IEEEauthorrefmark{12}
{ott@lrz.de},
\IEEEauthorrefmark{19}
{\{patki1,shoga1\}@llnl.gov},
\IEEEauthorrefmark{20}
{ivybopeng@kth.se},
\IEEEauthorrefmark{5}
{ssimms@lbl.gov},
\IEEEauthorrefmark{15}
{mung@illinois.edu},\\
\IEEEauthorrefmark{11}
{d.tiwari@northeastern.edu},
and
\IEEEauthorrefmark{9}{keiji.yamamoto@riken.jp}}

\maketitle

\begin{abstract}
Many High Performance Computing (HPC) facilities have developed and deployed
    frameworks in support of continuous monitoring and operational data
    analytics (MODA) to help improve efficiency and throughput. 
Because of the complexity and scale of systems and workflows and the need for low-latency response to address dynamic circumstances, automated feedback and response
have the potential to be more effective than current human-in-the-loop approaches which are laborious and error prone. 
Progress has been limited, however, by factors such as the lack of
infrastructure and feedback hooks, and successful deployment is often site- and case-specific. 
In this position paper we report on the outcomes and plans from a recent
    Dagstuhl Seminar, seeking to carve a path for community progress in the development of autonomous feedback loops for MODA, based on the established formalism of similar (\mapek{}) loops in 
autonomous computing and self-adaptive systems. 
By defining and developing such loops for significant cases experienced across HPC sites, we seek to extract commonalities and develop conventions that will 
facilitate interoperability and interchangeability with system hardware, software, and applications across different sites, and will 
motivate vendors and others to provide telemetry interfaces and feedback hooks to enable community development and pervasive deployment of MODA autonomy loops.
\end{abstract}


\input{1Introduction}
\input{2Approach}
\input{3UseCases}
\input{4DesignChallengesOpportunities}
\input{6Conclusion}
\input{7Acknowledgement}

\newpage
\bibliographystyle{IEEEtran}
\bibliography{WAFVR} 

\end{document}

%% file: 1Introduction.tex
\section{Introduction}
\label{s:introduction}
Using monitoring data to extract actionable insights on system behaviors
regarding facilities and building infrastructure, system hardware, system
software, and applications has been referred to as monitoring and
operational data analytics (MODA)~\cite{netti2021conceptual}.
Many high performance computing (HPC) and data centers have developed and/or deployed data collection frameworks that facilitate continuous and holistic monitoring and analysis (e.g., ~\cite{LDMSSC14,netti2019facility,9229641,10.1145/3149393.3149395,10.1007/978-3-031-23220-6_14,227665, 10.1145/3309205, bourassa2019operational, 5161234,osti_1819812, 10.1145/3225058.3225086, OperationalAnalyticsTrinity, ClusterCockpit}) in support of MODA goals.
While the \textit{monitoring} component of MODA is well established and deployed,
the \textit{analysis} part is still performed mostly by visual inspection, and feedback  typically involves a \emph{human in the loop}~\cite{9229641,10.1145/3149393.3149395,9355272} to make analysis-based responses.
As HPC systems are growing larger and ever more complex, manually analyzing high-dimensional time-series operational data becomes intractable 
without \emph{automation}. 
Moreover, having a human in the loop limits 
the speed of response and consequently, the opportunities for feedback-driven improvements. 

Some HPC and data centers have started to work on automated feedback and
response functionalities 
(e.g.,~\cite{10.1145/2907294.2907316,9926317,9820699,10.1007/978-3-031-23220-6_14,10027131,10.1145/3309205,ansel2014opentuner,9139855,227665,10.5555/3433701.3433715,
10.1145/3447818.3460362, 6968670 }). However, these are often one-off
approaches that are limited in the scope of data and responses and dependent on the specifics of particular use cases,
architectures, available actuators for responses, and so forth. 
It has become evident that closer collaboration would enable the community to better leverage aggregate work and experience to maximize impact.

\begin{figure}[htbp]
\centerline{\includegraphics[width=176.4pt]{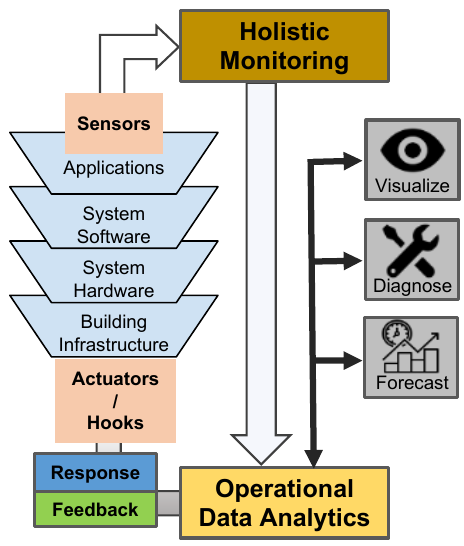}}
\caption{Vision of holistic monitoring and operational data analytics.}
\label{fig:holistic-moda}
\end{figure}

With this in mind, a cross-section of the HPC ecosystem gathered at Schloss Dagstuhl in April 2023 for a seminar on ``Driving HPC Operations With
Holistic Monitoring and Operational Data Analytics''~\cite{Dagstuhl23171}, with the goal of advancing the field and establishing a community path forward.
During the seminar, we converged on the high-level functionalities of \emph{monitoring}, \emph{analysis}, \emph{feedback}, and \emph{response} to comprise MODA-specific \emph{autonomy loops}, as illustrated in Fig.~\ref{fig:holistic-moda}. Then, a set of important use cases, across multiple sites, for autonomy loops were identified, which the group plans to develop as prototypes.
Through the prototypes, we intend to extract commonalities and develop
conventions that will facilitate interoperability with system hardware,
software, applications, etc., across different sites.
The use cases and established conventions will then be used to motivate vendors and others to provide the necessary telemetry and feedback hooks.

In this paper we posit that bringing formalism to MODA autonomy loops will help the community build more generalized interfaces, infrastructure, and interactivity approaches enabling reusable and more comprehensive (in terms of system,  applications, and facility components and response opportunities) approaches to self-adaptivity in HPC systems. 
We intend to leverage the formalism defined in the fields of autonomous
computing and self-adaptive systems~\cite{Weyns2020}, in the form 
of
\mapek{} loops: \monitor{}, \analyze{}, \plan{}, and \execute{} over some
\knowledge{}~\cite{kephart2003vision}, because of the similarity to our
concepts of monitoring, analysis, feedback, and response. 
This would both
enable a common context for thinking about the problem and allow us to take advantage of the \mapek{} architectural design patterns.
We provide background on the autonomy loop approach and 
considerations to help identify opportunities for developing generalizable \mapek{} loops-based infrastructure and approaches for MODA.
We highlight details for a foundational case that we intend to tackle as a community. 

%% file: 2Approach.tex
\section{Autonomy Loop Case Study Methodology}
\label{s:approach}

\begin{figure*}[htbp]
     \centering
     \begin{subfigure}[t]{0.14\textwidth}
         \centering
         \includegraphics[width=\textwidth]{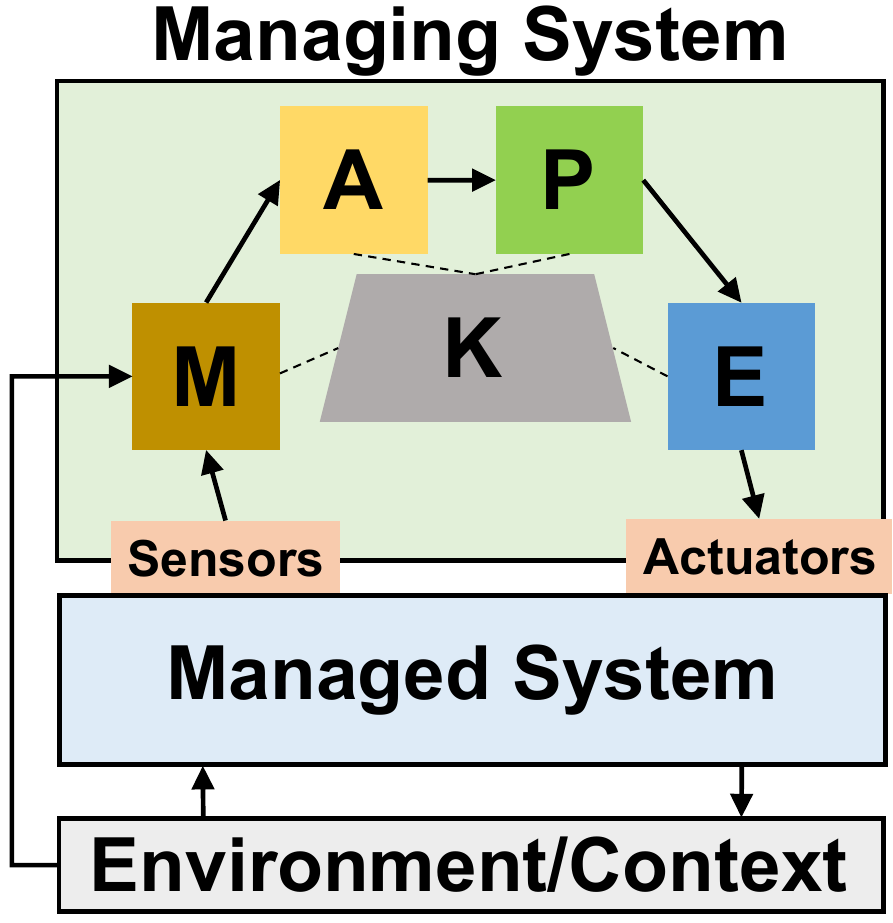}
         \caption{Classical MAPE-K loop}
         \label{fig:mape1}
     \end{subfigure}
     \hfill
     \begin{subfigure}[t]{0.28\textwidth}
         \centering
         \includegraphics[width=\textwidth]{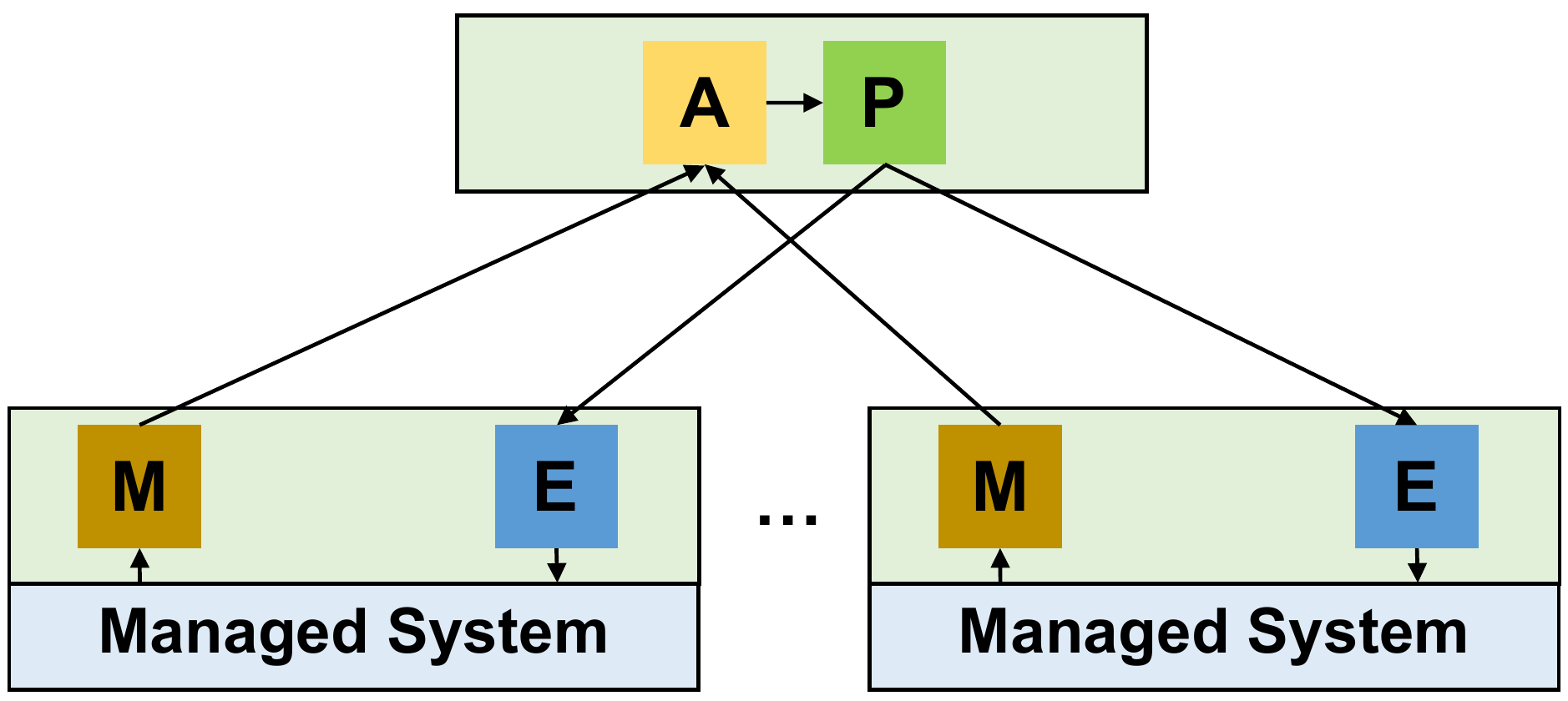}
         \caption{master-worker MAPE-K loop}
         \label{fig:mape2}
     \end{subfigure}
     \hfill
     \begin{subfigure}[t]{0.28\textwidth}
         \centering
         \includegraphics[width=\textwidth]{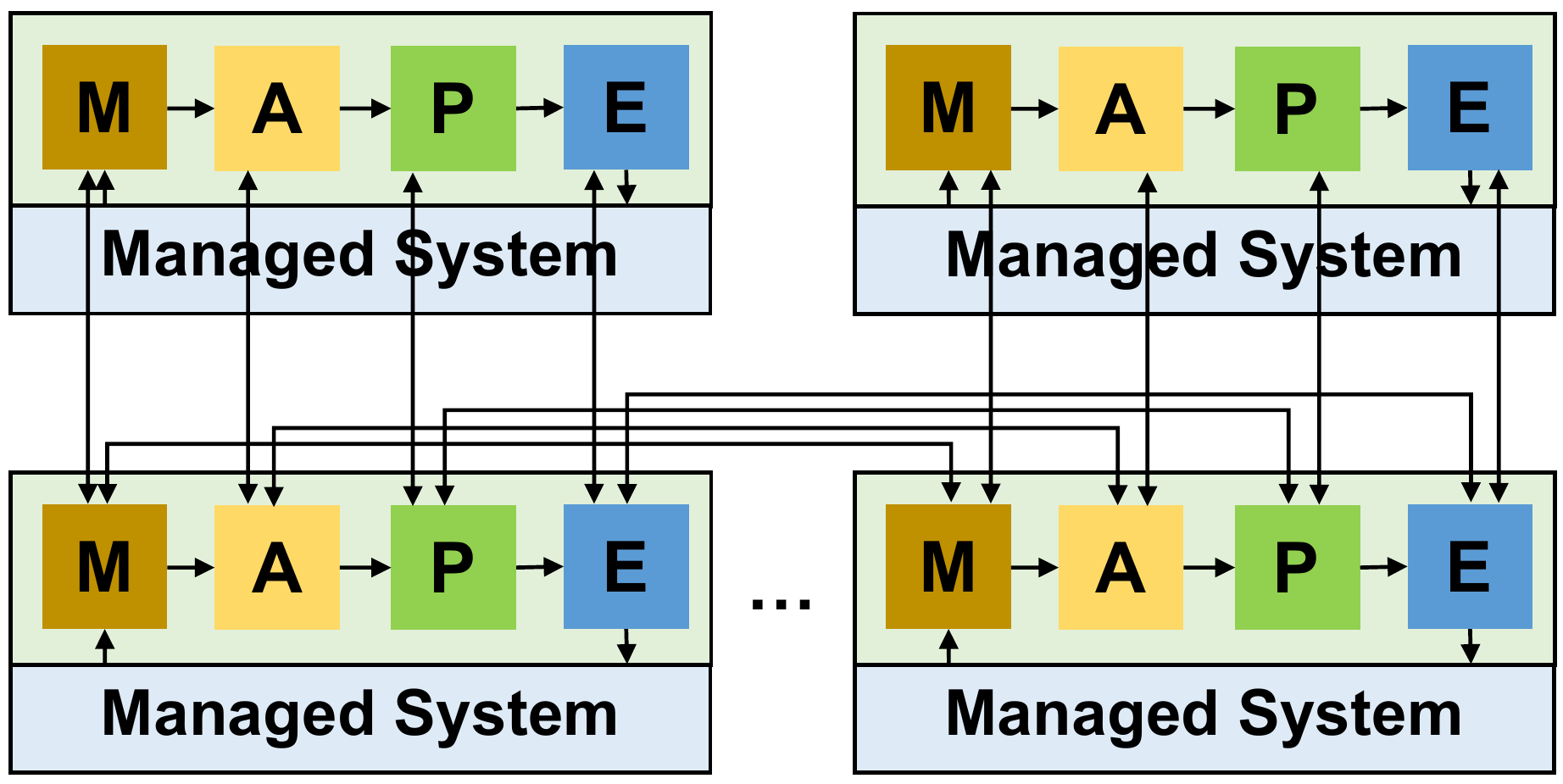}
         \caption{Fully decentralized, coordinated MAPE-K loop}
         \label{fig:mape3}
     \end{subfigure}
     \hfill
     \begin{subfigure}[t]{0.28\textwidth}
         \centering
         \includegraphics[width=\textwidth]{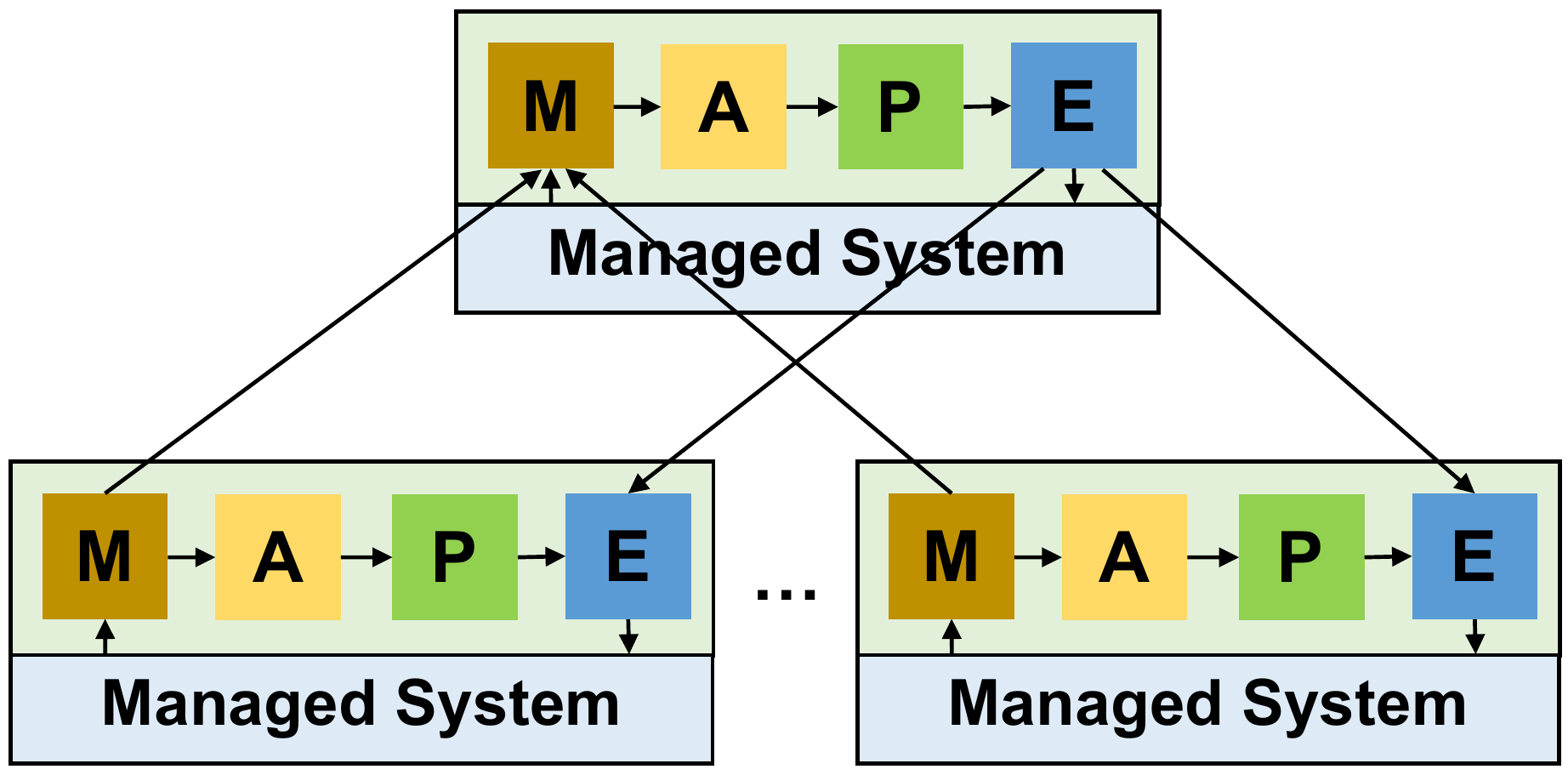}
         \caption{Hierarchical MAPE-K loop}
         \label{fig:mape4}
     \end{subfigure}
\caption{Design Patterns for \mapek{} loops. Leveraging the \mapek{} formalism will facilitate application of the designs to MODA autonomy loops.}
\label{fig:mapeoptions}
\end{figure*}

\mapek{} refers to the architectural pattern using \monitor{},
\analyze{}, \plan{}, and \execute{} loops over some \knowledge{}, introduced
in the autonomic computing (AC) initiative by IBM~\cite{kephart2003vision}
and used in self-adaptive systems~\cite{Weyns2020}.
The AC concept is based on constant checking and optimization of system status through adaptive decisions. 
The AC reference model includes a managed system, sensors, actuators, 
a managing system (utilizing \mape{}), and \knowledge{} about the managed system and its environment (where the latter is not under control).
In our context a \textit{managed system} is any HPC hardware or software system in which \textit{sensors} provide data about the HPC system, and \textit{actuators} are response hooks in building infrastructure, system hardware, system software, or applications. 
\monitor{} refers to the process of collecting data about an element of
interest, for example, an application.
\analyze{} and \plan{} refer to analyzing the collected data and planning an
appropriate response, for example, checkpointing.
\execute{} refers to carrying out the planned response through the use of response hooks.
\knowledge{} is pervasive in the components of the \mapek{} loop, as shown
in the ``classic'' (leftmost) loop in Fig.~\ref{fig:mapeoptions}. It can
include, for example, progress rate of an application compared with that of
a previous run, as well as knowledge gained from assessing the effectiveness of the \plan{} and \execute{} phases of previous loop iterations.

Arbitrarily complex autonomous actions can be supported by different
decentralized architectural design patterns for \mapek{}
loops~\cite{Weyns2013}, in which the \mape{} functionalities can be realized
by multiple components that coordinate with one another in different ways.
Some established patterns are illustrated in Fig.~\ref{fig:mapeoptions},
which shows, from left to right, 
the \textit{master-worker}, the
\textit{coordinated control}, and the \textit{hierarchical control}
patterns\footnote{We omit the \knowledge{} component and how it is used and
shared by the \mape{} components to focus on the interaction between \mape{}
loop components and across \mape{} loops and decrease the complexity. How
\knowledge{} can be stored and exchanged among MAPE components paves the way
for additional patterns.}. The first decentralizes only \monitor{} and
\execute{}; the centralized \plan{} can achieve global objectives and
guarantees but suffers from limited scalability, especially when managing a
complex system. The coordinated control pattern relies on fully
decentralized \mape{} loops that control different parts of the managed
system and have the potential of good scalability and robustness, but
decentralized \plan{} policies may suffer from instability and side-effects
due to indirect interactions. In the hierarchical control pattern,   decentralized \mape{} loops are organized in a hierarchy, with separation of concerns and time scales and aiming to improve scalability without compromising stability; however, division of control is not trivial. 

Previous works have demonstrated the potential for autonomy loops to improve efficiency in HPC operations.
Examples include throttling network or storage traffic in response to observed congestion or optimizing
cache policies in response to observed memory access patterns.
Progress has been limited, however, by available data, hooks, and opportunities for generality. 
To further such efforts, we seek to identify commonalities and conventions that could drive the development of widely reusable and interoperable infrastructure for MODA autonomy loops, leveraging the \mapek{} formalism and architectures.
A unified and generalized approach to solutions would simplify integration of autonomy loop components and functionalities, supporting a number
of complex subsystems, and would avoid an approach involving a confusing mix of disparate solutions. Considerations in determining designs for MODA cases would include access to monitored components and response options, response latency, desire for separability of functionality, need for coordination, and scalability and robustness. 

Beyond design and development, establishing autonomy loops within HPC software stacks and infrastructure will also depend on satisfying concerns of security, trust, and validity. We therefore 
propose the following five key questions be considered in 
approaching autonomy loop use cases to help identify opportunities for broader impact:
\begin{enumerate}[i)]
\item Can the autonomy loop be described in terms of high-level components with distinct responsibilities?
\item What interfaces or data formats would enable those components to be interchangeable?
\item What sort of open datasets would facilitate the use case?
\item How would validation be performed and user and system administrator trust be earned in order to enable autonomous actions?
\item How would practitioners be incentivized to engage and provide additional support (such as providing functionality and hooks for data, feedback, and responses) and usage of the loops in production?
\end{enumerate}

%% file: 3UseCases.tex
\section{Driving Generalization Through Initial Case}
\label{s:usecases}

We posit that a collaborative community approach to the development of MODA autonomy loops will further development, adoption, and production deployment.
To this end we have identified an initial set of specific use cases that could benefit from use of autonomy loops and that target scenarios
prevalent in production HPC. The diversity of cases is intended to
enable exploration of commonalities that could drive wider interoperability.
The cases are as follows:
\begin{enumerate}
\item \Maintenance{}: Responses to system maintenance events to ensure continuity of running jobs.
\item \IOQoS{}: Refinement of a storage system whose users receive QoS allocations through the use of \mapek{} loops of decreasing size and increasing automation, from rough estimates over a research campaign to parametric alteration based on profiling. The goal would be to adapt QoS parameters based on the current application performance and system I/O load to decrease interference, reduce tail latency, and provide more consistent results for deadline dependent workflows.
\item \OST{}: Response by an application, from continuous evaluation of
    storage back-end write performance, to close files using a poorly
        performing OST object storage target (OST), that is, a storage volume of a parallel filesystem such as Lustre). The application would then reopen them
using different OSTs, or explicitly request to avoid that OST in a case where the filesystem would allow it.
\item \Misconfiguration{}: Detection of misconfiguration of user jobs such
    as unintended mismatch of threads to cores, underutilization of CPUs or
        GPUs, or wrong library search paths. Depending on the type of misconfiguration, users could either be informed about their mistake along with suggestions for better configurations, or the misconfiguration could be corrected on the fly.
\item \Scheduler{}: Modification of a job's allocated run time based on
    continuous evaluation of its projected time to completion. This would
        also be extended to enable the scheduler to signal an application to checkpoint based on the time needed to write a checkpoint and the time remaining in an allocation.
\end{enumerate}

\begin{figure}[htbp]
\centerline{
\includegraphics[width=201.6pt]{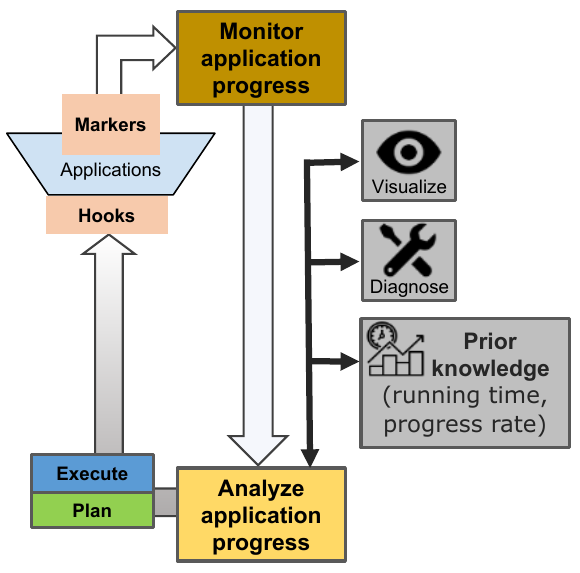}}
\caption{\Scheduler{} use case and its \mapek{} loop components.}
\label{f:schedulermape}
\end{figure}

We will initially focus on the \Scheduler{} case to define and develop our first set of common components. The \mapek{} autonomy loop is described here and illustrated in Fig.~\ref{f:schedulermape}:
\begin{itemize}
\item \monitor{} progress of an application. This could be via markers that could be output by an application (e.g., simulation time-step) or via progress information based on function calls or any application-relevant convergence criterion.
\item \analyze{} the progress relative to representative historical
    application run times, which would need to be collected and stored along
        with appropriate metadata. Given an application, a strategy is also
        required to map the application to a set of measurements of behavioral characteristics to enable comparison against past and future runs.
\item \plan{} action to be taken. This should take into account prior \knowledge{} of running time and progress rate (which might have to be inferred from similar jobs with different input decks). This may also take into account system state and expected changes due to projected changes in workload and associated resource utilization.
\item \execute{} the determined response.
Although the determined response may be to inform an application that it needs to request a run 
time extension or even to make the request on behalf of an application, the scheduler may deny the request or provide a shorter extension than requested.
\item Assess the \knowledge{} about the success of the \plan{} and refine
    the \knowledge{} through subsequent \textit{Monitoring} of the job's progress, iterating the \mapek{} loop. Note that this needs awareness of 
    whether or not the request was honored by the scheduler.
\end{itemize}

The \Scheduler{} case was chosen as the initial case for several reasons. First, we can gain insight into the variation of progress markers and run time through experimentation. This data should be straightforward to obtain and will be foundational to assessing the potential value of the work and developing the logic for the \plan{}. 

Second, we can obtain reasonable initial functionality with a single ``classical'' autonomy loop per application with loosely coupled implementations needed for the interactions between \mapek{} components. Interactions necessary for the \monitor{} and \analyze{} phases could be simply done by having the application's rank 0 drop time-steps periodically to a file or memory region to then be used in the progress assessment. Thus the components performing the progress assessment and determining the action do not have to be in the same process space as the application. Also, the components performing the \execute{} phase for scheduler interaction only require the rights to invoke the scheduler to increase the wall time for the job. For typical HPC schedulers, such as SLURM~\cite{SLURM}, this is an existing command-line functionality.

Third, a few simple measurable quantities can be used to forecast time to completion which will be used, in conjunction with the remaining
allocation time, to plan what action, if any, to take. These same quantities
can also be used to assess the effectiveness of the \plan{} (e.g., over/under-estimation of a change to time allocation). 

Fourth, a clear path exists to explore extensibility of the design and interactions from the simple prototype. Including an option for invoking asynchronous checkpointing would drive design of increasingly complex \plan{} and \knowledge{} components and application interactions. 

Exploring extensibility will be foundational for developing our other cases as well. The following examples describe such relationships between the \Scheduler{} case and our other four target cases: 
1) the \Maintenance{} case would use equivalent application interaction as invoking asynchronous checkpointing, 
2) the \IOQoS{} case would utilize the same capability for storage/retrieval/comparison of behavioral attributes of an application (e.g., I/O bandwidth profile) along with application interaction with extension to provide guidance on appropriate times to perform I/O, 
3) the \OST{} case would again utilize the capability for storage/retrieval of behavioral attributes in order to have a reference for expected operation along with application interaction to inform another response, and 
4) the \Misconfiguration{} case would likewise require storage/retrieval of
behavioral attributes and relationships to compare an application run with
expectations along with application interaction and, potentially, scheduler interaction.

The development process for the \Scheduler{} use case will follow the questions in Section~\ref{s:approach}:

\begin{enumerate}[i)]
\item 
The monitoring system will \monitor{} the application progress, potentially
        with help from the runtime system or instrumentation of the
        application. A yet-to-be-developed service will 
\analyze{} the progress and \plan{} the intervention. The scheduler will
        \execute{} the run-time extension.
\item
The actual interfaces will be determined during development, but tools and
        frameworks already exist that could be leveraged. The relationships, listed above, between the \Scheduler{} and other cases will help determine possibilities for common interfaces and interoperability.
\item
We plan to release the exploratory datasets used to gain insight into the
        variation of progress markers and run-time variation as open
        datasets.
\item 
Validation of the run-time extension will be clear through comparison of the time extension with the actual application run time. 
Trust by users and system administrators would require that other workloads and jobs were not adversely impacted by the extension mechanism.
This could be done by additional controls, such as limits on the number and overall time of extensions for a single application, and evaluations such as run time overestimations that would  have resulted in untaken backfill opportunities.
\item
Adopting an autonomy loop that increases their jobs' execution success would incentivize users.
Additional statistics, such as increase in completed and decrease in resubmitted jobs, would incentivize administrators to deploy it. 
Success in the \Scheduler{} case could also motivate developers and users to implement hooks for a checkpointing response as well, since a job would not be extended indefinitely.
\end{enumerate}

%% file: 4DesignChallengesOpportunities.tex
\section{Design Changes for Autonomous MODA}
\label{s:design}

Autonomy loops are the ``killer case''  for the MODA community. The ability to continuously make and enact data-driven decisions without requiring a human in the loop motivates the collection, analysis, and retention of holistic data at higher fidelity than ever before. This will cause both changes in and opportunities for design strategies for MODA autonomy loops. 

Increases in core counts have long been seen as providing an opportunity to co-locate analytics closer to compute resources. However, this opportunity has not been widely realized. Our autonomy use cases target new and tighter interactions with applications and system software both for gathering information and enacting response. 
Ideally, an established standardized set of interfaces for each component type would make the components interchangeable and the loop(s) modular, however, different requirements and associated implementations (e.g., latency, sampling rates, cardinality, high availability for monitoring) may drive multiple interfaces and interactions.
Therefore, \textit{interoperability} and \textit{interchangeability} are key design considerations.
We will support these by ensuring well-defined and documented interfaces for interactions we develop.

Increasing possibilities for low-latency actions will provide more motivation for \emph{in situ} analytics and decision-making, and hence storage in MODA designs. This will also drive more complex \mapek{} design patterns than first investigated in the \Scheduler{} case. Distributed autonomy, where each component has some decision-making capability and decision-executing authority (as discussed in Section~\ref{s:approach}), 
will be useful for robust and resilient operations. Agent-based models have shown this in the context of distributed systems~\cite{agent}. Resilience is essential in HPC systems where operations must persist through component and subsystem failures.

Failure prediction and anomaly detection have long been MODA analysis goals. Our cases further analytics in continuous performance characterizations and comparisons. Our storage architecture decisions will then increasingly consider metadata representations for models, moving beyond traditional considerations of insert rates for raw time-series data. 

Relatedly, our analyses will also be expanded to include determination of
confidence in the models for decision-making and assessment of the
effectiveness of the \plan{} and \execute{} phases. Confidence measures are
required as we move beyond human-in-the-loop decision-making, 
particularly for safe operations of power and energy controls. 

Note that autonomy loops in HPC operations do not have to replace the human-in-the-loop approach, and could complement it. A \emph{human-\textbf{on}-the-loop} approach would have the loop continue without waiting for user and administrator input, but sending them notifications and explanation about decisions that allow for observing its effects when necessary~\cite{Li_seams2020}. The decision-making would then also include execution of contingency plans for when the humans are absent.

The HPC domain should look to AI tools, algorithms, and generated models to help drive the automated decision process.
However, focus should be on careful selection of efficient models and modeling parameters that fit HPC data.
For instance, the present outlook in the AI community is the use of large
models with millions of parameters. However, such models may not be 
efficient when complex optimizations for real-time decisions must be made. In fact, the constantly evolving nature of the environment requires continual/lifelong AI that can evolve rapidly with small overhead~\cite{Gheibi_seams22}. Moreover, precision requirements must be built into these AI models, and the lack of \textit{interpretability} and \textit{explainability} must be addressed to get robust AI models. Furthermore, use of AI for autonomous loops may impact resources allocated to applications. 
In summary, simply applying the present AI tools and algorithms will not be sufficient and ample opportunities exist for further design and development.

%% file: 6Conclusion.tex
\section{Conclusion}
\label{s:conclusion}

In this position paper we have reported on the outcomes and paths forward
        from a recent Dagstuhl Seminar, seeking to carve a path for
        community progress on the development of autonomous feedback loops
        for MODA, based on the established formalism and architecture of
        \mapek{} loops in autonomous computing and self-adaptive systems.

We are presenting our position early in our work process in order 
to get input from the wider HPC community and to engage collaborators interested in determining interfaces for HPC autonomy loop components and developing interoperable and interchangeable components that utilize those interfaces.

Additionally, we seek to facilitate development by encouraging testbeds to be defined and made available to the community. The main obstacle to exploration of autonomy loops is the fear of potentially intrusive changes to the system behavior, often deemed unacceptable on production systems. Nevertheless, it may be possible to utilize stranded resources of HPC systems that are being decommissioned for a limited time, or include experiments during bring-up or extended maintenance. However, we believe that the best approach is to include MODA targets into the system definition itself. Such efforts will be propelled by well-defined modularization of \mapek{} components and their associated APIs, so that individual components can be replaced while preserving appropriate system boundaries, enabling appropriate auditing and trust levels.
The OpenCUBE project~\cite{opencube} is aiming to provide such opportunities, by defining a process to submit \mapek{} loop experiments to be executed on their testbed system.

%% file: 7Acknowledgement.tex
\section*{Acknowledgment}
\label{s:acknowledgement}

The authors thank Schloss Dagstuhl for hosting Seminar 23171 ``Driving HPC
Operations With Holistic Monitoring and Operational Data Analytics.''

The work is jointly supported by the European Union's Horizon 2020 research and innovation programme (grant agreement No. 957407, DAPHNE).

This work was performed under the auspices of the U.S. Department of Energy by Lawrence Livermore National Laboratory under Contract DE-AC52-07NA27344 (LLNL-CONF-851925).

Sandia National Laboratories is a multimission laboratory
managed and operated by National Technology \& Engineering
Solutions of Sandia, LLC, a wholly owned subsidiary of Honeywell International Inc., for the U.S. Department of Energy’s
National Nuclear Security Administration under contract DE-
NA0003525. This paper describes objective technical results
and analysis. Any subjective views or opinions that might
be expressed in the paper do not necessarily represent the
views of the U.S. Department of Energy or the United States
Government.

This work was supported by the U.S. Department of Energy, Office of Science, Advanced Scientific Computing Research, under Contract DE-AC02-06CH11357.

This work was funded by the European Union under the Horizon Europe program's OpenCUBE project, grant agreement 101092984.